\documentclass[12pt]{article}
\usepackage{graphicx}
\usepackage{amssymb,amsmath,amsfonts,palatino,amsthm}
\usepackage{amssymb}
\usepackage{epstopdf}
\newcommand{\f}{\begin{equation}}
\newcommand{\ff}{\end{equation}}

\begin{document}

\title{Non-local beables\\}
\author{Lee Smolin\thanks{lsmolin@perimeterinstitute.ca} 
\\
\\
Perimeter Institute for Theoretical Physics,\\
31 Caroline Street North, Waterloo, Ontario N2J 2Y5, Canada \\
and\\
Department of Physics and Astronomy, University of Waterloo\\
and\\
Department of Philosophy, University of Toronto}
\date{\today}
\maketitle

\begin{abstract}

I discuss the idea that the beables underlying quantum physics are non-local and relational, and give an 
example of a dynamical theory of such beables based on a matrix model, which is the bosonic sector of the $BFSS$ model.  Given that the same model has been proposed as a description of $\cal M$ theory, this shows that quantum mechanics may be emergent from a theory of gravity from which space is also emergent.

Submission to the John Bell Workshop 2014, of the International Journal of 
Quantum Foundations.   
\end{abstract}

\tableofcontents

\newpage

\section{Taking non-locality seriously}

I would like to begin with a remark of John Bell on the possibility that the beables are non-local.

{\it ``Of course, we may be obliged to develop theories in which there are no
strictly local beables. That possibility will not be considered here\cite{Bell-nlb}."}

When I read that, I was astounded because it made me realize that ever since encountering Bell's theorem as a first year undergraduate I have assumed that there are non-local beables; indeed most of my work in quantum foundations has been a search for them.  The reasons to expect the beables are non-local are easy to state.

\begin{itemize}

\item{}{\bf Non-locality in quantum gravity.}  If the metric of space-time is a quantum operator subject to quantum fluctuations then locality must be only a feature of the  classical approximation.  Non-locality must arise as a consequence of quantum fluctuations of the metric.  And these cannot be limited to the Planck scale; there are several arguments that show that non-locality must be present in quantum gravity at large scales.  Some of these come from attempts to solve the black hole information paradox
(black hole complementarity, EPR/ERB duality), others come from the ubiquity of defects in locality in non-perturbative treatments of quantum gravity\cite{dl}.  

\item{}{\bf Relationalism.}  Basic to the thinking of many of us in quantum gravity is the thesis of relationalism, that holds that the fundamental beables describe relationships among elementary events or particles.  That is, the hidden variables do not give a more detailed description of the inner workings of an electron, they describe details of relations between the diverse electrons in the universe that are ignored under the coarse graining that gives rise to the emergence of space.  These can be called {\it relational hidden variables.}   

\item{}{\bf Space is emergent.} One thing the diverse approaches to quantum gravity agree with is that space is not fundamental, but emergent.  More fundamental than space is a network of relations,  which constitute the basic ontology of the theory.  This more fundamental and relational network of relations has been described as a graph (loop quantum gravity, quantum graphity), a matrix (string theory), a partial order (causal set theory), a dual triangulation ( causal dynamical triangulations and spin foams), but what all these have in common is the hypothesis that space is not part of the basic ontology of the world.      But if space is emergent, so is locality.  This suggests that the non-locality of quantum theory is described by beables that are ordinary beables at the non-local (or better: {\bf a-local}) level that become part of the quantum state when space emerges.  In other words, space and the quantum state emerge together, each carrying part of the information in the fundamental non-local ontology.  

\end{itemize}

This leads to a hypothesis.  {\it  The fundamental beables are relational and a-local, having their fundamental description in a phase from which space has yet to emerge.  Space and quantum theory emerge at the same time.  The stochasticity of quantum theory arises from our lacking control over beables that describe relationships between a system and other, distant systems in the universe.}

\section{A non-local hidden variables theory}

Can the hypothesis just stated be expressed in a detailed dynamical theory of relational hidden variables, from which quantum mechanics can be derived?  Yes, and it has been done several different ways\cite{mehidden1,thispaper,mehidden2,Adler,Artem}.

Here is a sketch of one way, which is described in detail in \cite{thispaper}, from which the following is taken.  

The beables of the theory are 
$d$, $N \times N$ real symmetric
matrices $X_{a i}^{\ j}$, with $a=1,...,d$ and $i,j=1,...,N$.  
The classical, local observables are taken to be the eigenvalues of these matrices, $\lambda^a_i$.  
These can be imagined to give the positions of $N$ particles in $d$ dimensional space.  
Relative to these, the matrix elements are non-local, as a shift in the value of any one matrix element perturbs all the eigenvalues.  Our aim is to give a dynamics to the matrices such that 
quantum dynamics emerges for their eigenvalues.

The
dynamics of these matrices is given by an action\footnote{This action is found in string theory, where it is  called the $BFSS$ matrix model, it also can be understood to arise from an $SO(N)$ Yang-Mills theory in an approximation from which spatial derivatives can be neglected \cite{CH,dWHN,BFSS,IKKT}.}, 
\f 
S= { \mu} \int dt Tr \left [
\dot{X}^2_a - \omega^2 [X_a,X_b][X^a,X^b]   \right ]
\label{action} 
\ff 
We choose the matrices $X^a$ to be
dimensionless. $\omega$ is a frequency and $\mu$ has dimensions of
$\mbox{mass}\cdot \mbox{length}^2$.  We do not assume $\hbar =1$,
in fact, as we aim to derive quantum mechanics from a more
fundamental theory, $\hbar$ is not yet meaningful.  We will
introduce $\hbar$ as a function of the parameters of the theory
when we derive the Schroedinger equation as
an approximate evolution law.
We may note that the parameters of the theory define an energy
$\epsilon = \mu \omega^2$.

The basic idea is that the matrix elements of $X^a$
will be the non-local hidden variables.    The theory is invariant under $SO(N)$ transformations,
\f
X^a \rightarrow U X^a U^T 
\ff
where $U \in SO(N)$.   These constitute gauge transformations, so the physical observables will be invariants under $SO(N)$.  These include the eigenvalues of the matrices
\f
\lambda^a_i
\ff

We note that the model has a
translation symmetry given by 
\f 
X^a \rightarrow X^a + v^a I.
\label{translations} 
\ff 
The result is that the center of mass
momentum of the system is conserved.

Now the potential energy 
\f 
U= { \mu  \omega^2}Tr \left [
 [X_a,X_b][X^a,X^b]   \right ]
\label{U} 
\ff 
has its minima when the $d$ matrices commute with each other, in which case they can be simultaneously diagonalized,
\f
X^a = D^a =  \mbox{diag}(\lambda^a_1. \ldots ,  \lambda^a_N  )
\ff
This will give the classical approximation, hence we take the eigenvalues to label the positions of $N$ identical particles in $R^d$.  At the classical level the N particles are free; but if we wanted to model a system with classical interactions we could add to the potential energy a function of the eigenvalues.  
\f 
S^\prime= { \mu} \int dt Tr \left [
\dot{X}^2_a - \omega^2 [X_a,X_b][X^a,X^b]   -V(\lambda )  \right ]
\label{action2} 
\ff 
To get quantum behaviour, we will put the system at a small, but finite
temperature, the result of which will be that the matrix elements
undergo Brownian motion as they oscillate in the potential.  It
follows from linear algebra that the eigenvalues also undergo
Brownian motion.  The work is then to show that the parameters of the theory
can be scaled with $N$ in such a way that 
quantum dynamics is realized for the eigenvalues.

I won't give the details here, but they can be found in \cite{thispaper}.  The key steps are as follows.

\begin{enumerate}

\item{} When the system is at finite temperature the eigenvalues can be shown to undergo Brownian motion.
Using the language appropriate to Brownian motion, which are stochastic differential equations, we derive a description of this Brownian evolution.  These are defined in terms of a probability density for the eigenvalues, $\rho (\lambda , t)$ and a probability current velocity, $v^a_i (\lambda, t)$.  These are related by probability current conservation,
\f
\dot{\rho}(\lambda , t) =  \frac{\delta \rho v^a_i (\lambda, t)}{\delta \lambda^a_i }
\ff

\item{}One shows that to leading order in $1/N$ the current velocity is irrotational, so it is the gradient of a scalar potential, $S(\lambda , t)$.
to leading order in $1/N$,
\f
\mu v^a_i = { \delta S_\lambda (\lambda ) \over \delta
\lambda_a^i} + O(1/N) .
\label{verygood}
\ff

\item{}We take advantage of a fundamental insight of Nelson\cite{Nelson} who showed that quantum mechanics can be understood as a form of Brownian motion with the unusual property that it is conservative.  That is, the motion is stochastic, but unlike the normal case in which Brownian motion is accompanied by dissipation, there is an average conserved energy.  This means the stochastic evolution of the eigenvalues becomes time reversal invariant.   To achieve this we consider the diffusion of the eigenvalues in this system.  The diffusion has two sources, thermal diffusion, related to being at a finite temperature, $T$, and large $N$ effects, coming from the fact that all matrix elements contribute a little bit to the motion of an eigenvalue.

We find conservative Brownian motion in a regime where these two effects are balanced.  As stressed in \cite{tuning} some kind of tuning is necessary to derive quantum mechanics, which is conservative and invariant under time reversal, from a general theory of stochastic motion, in order to cancel dissipative effects.

To describe this regime we study the stochastic dynamics of the eigenvalues in a particular limit where we take 
the size of the matrices, $N \rightarrow \infty$ while we take the temperature $T \rightarrow 0$ in a way that keeps a fixed value for the diffusion constant $\nu_\lambda$ fixed.

This regime is defined by holding fixed a dimensionless scaled temperature
\f 
t = { N T \over 8 (d-1) \mu \omega^2} 
\ff 

The diffusion constant for the eigenvalues 
in this limit is  shown to be 
given by
\f
\nu_\lambda = \omega  \frac{d t^{3/2}}{4(d-1)^{3/2}} 
\label{nulambda}
\ff

\item{}
We then define the wave functional
\f
\Psi (\lambda , t ) = \sqrt{\rho (\lambda , t ) } e^{\imath S(\lambda , t )/ \hbar }
\ff
where Planck's constant is defined by
\f
\hbar = \mu \nu_\lambda = \mu \omega {t^{3/2} d \over 4(d-1)^{3/2}  }
\ff

\item{}There is one last step which involves subtracting out a certain divergent energy
$E_Q^\prime$.
We use this to renormalize the wavefunctional
so that 
\f 
\Psi_r (\lambda ) = e^{iE_Q^\prime t /\hbar} \Psi
(\lambda ) 
\ff 
This we are able to show satisfies the free Schroedinger equation,
\f
\imath \hbar {d
\Psi_r(\lambda, t)  \over dt} = \left [ -{ \hbar^2 \over 2 \mu}
{\delta^2 \over \delta ( \lambda^a_i)^2} \right ] \Psi_r
(\lambda, t)
\ff

\item{}If we added interactions $V(\lambda )$ we find instead
\f
\imath \hbar {d
\Psi_r(\lambda, t)  \over dt} = \left [ -{ \hbar^2 \over 2 \mu}
{\delta^2 \over \delta ( \lambda^a_i)^2}  + V(\lambda ) \right ] \Psi_r
(\lambda, t)
\ff
\end{enumerate}

\section{Implications}

The model I have sketched shows that quantum mechanics can be recovered from an explicit hidden variables model whose beables are non-local.  This is in accord with the reasons I stressed that the beables of quantum theory should be taken as non-local.  I would thus propose that the ultimate legacy of Bell's fundamental work will be the discovery  that quantum theory is a description of an a-local world, which we happen to see in a phase where space has emerged.  When we try to describe the physics of local subsystems of the universe, delineated by the emergent and approximate concept of locality, we are forced to neglect interactions which are really there between the subsystem's microscopic degrees of freedom and other degrees of freedom now emerged in distant parts of the universe.  These non-local interactions are mediated by relational degrees of freedom that are non-local, in the sense that they are shared between subsystems that are distant from each other in the emergent concept of locality.

Because of the neglect of these non-local degrees of freedom, the quantum physics of local subsystems is stochastic and subject to a persistent and universal Brownian motion, which is the cheshire cat smile of the fundamental a-locality of the world.  In this sense $\hbar$ is a measure of the resistance of the world to a local description.  


\section*{ACKNOWLEDGEMENTS}

I would like to thank Dr. Shan Gao for the encouragement to contribute this paper.

This research was supported in part by Perimeter Institute for Theoretical Physics. Research at Perimeter Institute is supported by the Government of Canada through Industry Canada and by the Province of Ontario through the Ministry of Research and Innovation. This research was also partly supported by grants from NSERC, FQXi and the John Templeton Foundation.

\end{document}